\documentclass[a4paper,11pt]{article}
\pdfoutput=1 

\usepackage{jinstpub} 

\usepackage{lineno}
\notoc{\notoctrue}

\title{Design and Implementation of a Compact Analog Constant Fraction Discriminator for High-Resolution Timing in Gamma-Ray Spectroscopy}

\author{Michael Wiebusch}
\affiliation{GSI Helmholtzzentrum für Schwerionenforschung,\\Planckstr. 1, 64291 Darmstadt, Germany}

\emailAdd{m.wiebusch@gsi.de}

\abstract{This work presents a custom analog front-end card designed for the read-out of PMTs coupled to lanthanum bromide scintillators.
It integrates 16 discrete analog constant fraction discriminators (CFDs) on a compact 12x10 cm board, providing precise timing information for nuclear lifetime measurements.
The design emphasizes cost-effectiveness, utilizing off-the-shelf discrete components, as well as compactness, achieved by using miniature coaxial connectors and cables as delay elements.
The focus of this paper lies on the unusual and extremely minimalist analog shaper/discriminator design which is devised without operational amplifiers, making use of only a handful of RF transistors and LVDS receivers in place of comparators.
}


\proceeding{
TWEPP 2024 Topical Workshop on Electronics for Particle Physics\\
September 30, 2024 to October 4, 2024\\
Glasgow, Scotland}

\begin{document}
\maketitle
\flushbottom

\section{Introduction}
\label{sec:CFD_theory}

\begin{figure}
\centering
\includegraphics[width=0.5\textwidth]{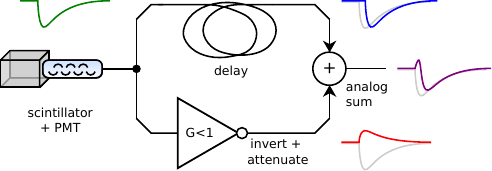}
\caption{The block schema of an ideal Constant Fraction shaper.}
\label{fig:ideal_CFD}
\end{figure}

A Constant Fraction Discriminator (CFD) is used to detect fast signals with precise timing, regardless of signal amplitude. 
Its main purpose is to counteract the so-called time-walk effect, i.e. an amplitude-dependent systematic deviation of the measured time-of-arrival, which affects simpler leading-edge discriminators. 
The key principle is to trigger based on a fixed fraction of the signal's peak amplitude, rather than the signal's leading edge.
All CFDs comprise the following (or equivalent) signal processing steps (see Figure~\ref{fig:ideal_CFD}):
\begin{enumerate}
\item
Splitting the incoming signal into two identical copies.
\item
One copy is delayed, e.g. by traveling through some length of coaxial cable.
\item
One copy is inverted and attenuated by a defined factor.
\item
Analog summation of both branches, resulting in a pulse waveform with a prominent zero-crossing.
\item
Detecting the zero-crossing yields the unskewed time-of-arrival information since the zero-crossings stay in the same relative position regardless of the input amplitude.
\end{enumerate}
The defining parameters of a CFD are the delay time and attenuation factor, both of which influence the
effective fraction of the pulse amplitude the device triggers on. Usually the best performance is achieved when the delay is of similar length as the rise time of the detector signal.
Originally devised as an analog circuit in the 1960s\cite{Gedcke}, the CFD is a more general concept that can also be implented as a digital signal processing algorithm. Both, the circuit version and the algorithm are actively used today\cite{jolie}.

Typical analog CFD circuits utilize multiple high-bandwidth operational amplifiers.
Such devices must be able to operate up to $100\,MHz$ or more\footnote{The amplifier actually has to have a gain bandwidth product far greater than $100\,MHz$ if significant gain has to be achieved at this frequency.}, to reproduce the steep leading edge (few $ns$) of the detector signal.
Suitable amplifier ICs tend to be costly and consume significant amounts of power.
A specialty of the design proposed in this paper, is the complete lack of op-amps - to save power, costs and board space. Instead it employs only a handful of low-cost RF transistors in combination with differential LVDS receivers in place of analog comparators.
Without sacrificing performance the heart of the circuit, the constant fraction shaper, was reduced to its most minimalist form, comprising only a single transistor and a delay line, apart from passive components.
The tradeoff for the achieved simplicity and compactness is its fixed use case, i.e.
the delay time and attenuation factor are optimized for the pulse shape of one particular detector.
The following circuit designs have been conceived for the read-out of $LaBr_3 (Ce)$ scintillators coupled to photomultiplier tubes (PMT) producing negative pulses with a $5.8\,ns$ leading edge time constant and a tail of $20\,ns$. However, the circuit concept should be applicable to broader range of detector types.


\section{A Minimalist Constant Fraction Shaper}
\label{sec:CFshaper}

\begin{figure}
\centering
\includegraphics[width=.8\textwidth]{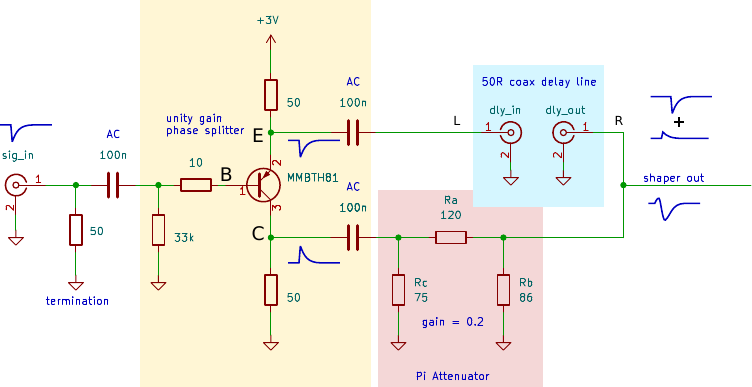}
\caption{The proposed circuit for a minimalist Constant Fraction shaper. This particular configuration of the Pi attenuator was calculated for an attenuation fraction of $F=0.2$.}
\label{fig:mini_CFD}
\end{figure}

Figure~\ref{fig:mini_CFD} shows the complete CF shaper consisting of only a transistor, a delay line, and a handful of passive components.
We can identify three main components of the shaper: The transistor wired as a unity gain phase splitter, a $50\,\Omega$ coaxial cable used as a delay line and a passive Pi-type attenuator network comprising the resistors $R_a, R_b$ and $R_c$. 
The phase splitter receives the detector pulse as a voltage signal across the termination resistor and creates two copies of the incoming signal: A direct copy at the emitter~(E) and an inverted copy at its collector~(C).\footnote
{Note that the less common PNP bipolar transistor is chosen over the NPN type, since it provides a wider dynamic range for strictly negative input pulses.}
The direct copy is AC-coupled into the transmission line while the inverted copy is AC-coupled into the Pi attenuator.
So far we have ingredients 1~to~3 from section~\ref{sec:CFD_theory}, i.e. splitting, creating a delayed copy and attenuated inverted copy.
Step 4, i.e. the analog summation is achieved by feeding the output of the attenuator into the right end of the delay line, which is also where we tap the shaper output signal.
Driving the line from both sides may seem confusing but has the following advantage:
The analog summation of the delayed and the inverted/attenuated signal takes place due to the fact that on a transmission line, waves which travel in different directions do not interact but simply superimpose.
This principle holds everywhere along the transmission line - even at the very end, as long as the end is terminated with a resistor equal to the wave impedance $Z_d$ of the line.
The Pi attenuator allows for impedance matching on both sides separately while (within limits) providing an additional degree of freedom for the attenuation fraction $F$.
Knowing this, we use the Pi network to \emph{also} terminate the delay line.
We have yet to determine the correct values for the three unknowns $R_a, R_b$ and $R_c$.
This is achieved by solving the following three-equation system:\footnote
{The "parallel" symbol implies the combined impedance of two components connected in parallel: $ A || B = Z_{A||B} = \left(A^{-1}+B^{-1}\right)^{-1}$}
\begin{eqnarray}
R_a || R_b & = & Z_{d} \label{eq1}\\
R_c || \left( R_a + \left(R_b || Z_{d} \right) \right) & = & Z_{d} \label{eq2}\\
\frac{R_b || Z_{d}}{R_a + \left(R_b || Z_{d}\right)} & = & F \label{eq3}
\end{eqnarray}
Equation~\ref{eq1} demands terminating the right side of the delay line with a resistance equal to its impedance $Z_{d} = 50\,\Omega$.
For simplicity we're approximating the impedance at the collector to be zero.
Eq.~\ref{eq2} demands that the collector experiences the same load as the emitter, which is again $Z_d=50\,\Omega$.
Only with matching loads the phase splitter will produce mirror copies of equal amplitude.
Finally eq.~\ref{eq3} describes the fraction $F$ of the inverted signal that is coupled into the right side of the delay line.
The general solution of the system of equations can be found as follows:
\begin{eqnarray}
R_a = \frac{Z_d}{2F},\quad R_b = \frac{Z_d}{1-2F},\quad R_c = \frac{Z_d}{2F^2-2F+1}
\end{eqnarray}
In the above calculation we assume that the circuit loading the shaper output has a relatively high input impedance ($\gg50\,\Omega$).\footnote{If this is not the case, the impedance of the next circuit stage has to be added into the equation system in parallel with $R_b$.}

\begin{figure}
\centering
\includegraphics[width=0.8\textwidth]{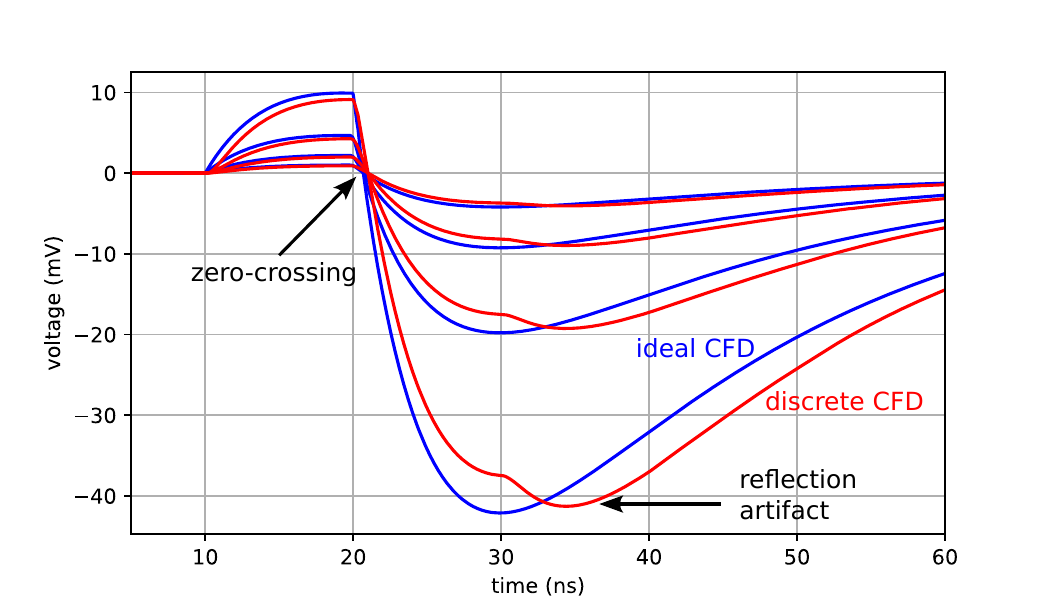}
\caption{Comparison of the simulated waveforms of an ideal Constant Fraction shaper (blue) and the proposed discrete shaper (red) for four different input amplitudes. Attenuator fraction $F=0.2$, delay time $=10\,ns$ (exaggerated for illustrative purposes).}
\label{fig:ideal_vs_real}
\end{figure}

Figure~\ref{fig:ideal_vs_real} shows the simulation of both, an ideal CF shaper and the proposed circuit side by side.
It is important to note that we lose (almost) no gain compared to the ideal CF shaper, even though the voltage summation is performed by passive components only.
The circuit has an additional feature manifesting as an extra bump near the maximum of the shaped pulse.
This artifact arises from driving the delay line from both sides but only providing proper termination on the right side.
The attenuated inverted signal travels back to the impedance mismatch at the collector ($Z_C\ll Z_d$)
where it gets inverted and reflected back into the delay line before it too is dissipated by the Pi network.
While this artifact deviates from the ideal pulse shape, it always occurs after the zero-crossing and is therefore not relevant to the operation of the CFD.

\section{The Zero-Crossing Detector}
\label{sec:ZCdet}

\begin{figure}
\centering
\includegraphics[width=0.8\textwidth]{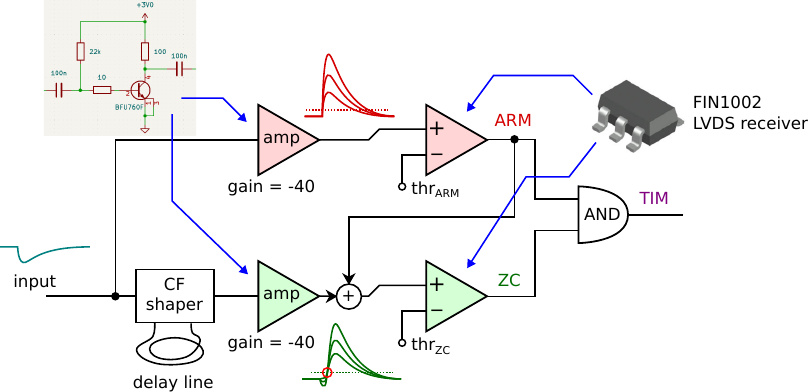}
\caption{Block schema of the proposed zero-crossing detector circuit.}
\label{fig:block_schema_CFD}
\end{figure}

So far we only discussed the constant fraction shaper, leaving out the final step of detecting the zero-crossing.
For this purpose we use inexpensive LVDS receiver ICs as fast comparators.
Figure~\ref{fig:block_schema_CFD} shows the additional circuitry required around the CF shaper.
The shaped pulse signal is fed into the positive input of the receiver
while the negative input receives the reference threshold DC voltage from a DAC.
The targeted input amplitude range spans from $10\,mV$ to $1\,V$, but
for the comparator to reliably react to the smallest pulses we require a gain stage in between the CF shaper and the comparator.
In this case, the necessary amplification is achieved though a simple common-emitter amplifier built with a single RF transistor (BFU760), which provides a gain of $-40$ or $32\,dB$.
With such a high gain, the amplifier is bound to clip and distort for all but the smallest input signals. Nevertheless, this is acceptable since the only relevant feature of the shaped signal is the zero-crossing (ZC) which is at the same level as the signal baseline.

A problem occurs when setting the comparator threshold directly at the baseline, leading to uncontrolled toggling of the comparator output most of the time.
To mitigate this, we employ a second comparator (with its own gain block) as an arming circuit which reacts to the unshaped raw input signal. The arming comparator has a threshold set just above the noise of the baseline and its output is used to gate the zero-crossing comparator by means of an AND gate.
Even with its output gated, the noise receiving ZC comparator still creates unwanted electromagnetic interference on the analog front-end.
This scenario is avoided with another trick: A small fraction (circa $150\,mV$) of the arming comparator's LVCMOS output is passively added to the CF-shaped and amplified signal.
Consequently, with the additional kick, the zero-crossing is quickly raised above the baseline, and the ZC comparator threshold can be set a safe amount above the baseline noise.

Even with a low threshold the arming comparator cannot instantly detect an incoming pulse, since it is always subject to the time-walk effect.
Nevertheless, the concept works as long as the "arming kick" occurs while the CF shaper is still in the undershoot phase leading up to the zero-crossing.
To maximise this margin, the propagation delay of the LVDS receiver plays an important role. For this reason the receiver IC FIN1002M5X was chosen over the more standard SN65LVDS2, since it was experimentally found to be a whole $1\,ns$ faster ($1.5\,ns$ vs $2.5\,ns$), while also being significantly cheaper.

\section{Implementation and Performance}
\label{sec:implementation}
The above circuit concepts were applied in the design of an analog front-end card, named "Twin\_Peaks\_CFD1",
tailored specifically to the needs of the Fast Timing Array (FATIMA), which is part of the DEcay SPECtroscopy (DESPEC)\cite{mistry} experiment at GSI/FAIR, Darmstadt.
FATIMA comprises 36 cerium-doped lanthanum bromide scintillators, coupled to traditional photomultiplier tubes, aimed at measuring lifetimes of exotic nuclear states by detecting gamma rays\cite{fatima}.
The necessity for excellent time-of-arrival information, coupled with clear spectroscopic distinction between gamma energies, led to the development of Twin\_Peaks\_CFD1, which hosts 16 discrete analog constant fraction discriminators (CFDs) on a compact $12\times10\,cm$ board.
The board interfaces seamlessly with the TAMEX4 Time-to-Digital Converter (TDC) board, a widely used FPGA-based TDC at GSI, offering precise time measurements with a resolution of $15\,ps$ rms.
Apart from the minimalist circuit concept, Twin\_Peaks\_CFD1's compactness is achieved through the use of 16x $50\,cm$ long, $1.3\,mm$ thick coaxial cables, each providing a delay of $2.5\,ns$.
The cables are coupled to the board via U.FL miniature coax connectors typically found in WiFi antenna assemblies. 
When neatly curled up and fixed with cable ties, the cables do not extend beyond the board edges, while the total thickness of the assembly remains below the height of the stacked coaxial connectors on the front edge.
CFD performance tests with a PMT-like test pulse yield a leading edge precision of $60\,ps$ (stdev)\footnote{Without correcting the residual walk. With an additional offline walk correction a precision of $30\,ps$ is possible. The performance is ultimately limited by the jitter of routing the CFD output signals through the on-board FPGA.} while the amplitude was varied up to factor 10 (20mV-200mV). Gamma-gamma coincidences with the CFD plus actual detectors yield a precision of better than $190\,ps$ (stdev).


\section{Conclusion}
\label{sec:conclusion}

The proposed minimalist Constant Fraction Discriminator design successfully achieves precise time-of-arrival measurement with a compact and cost-effective solution, built entirely without operational amplifiers.
By utilizing an unusual, extremely stripped down configuration of RF transistors, miniature cable assemblies and differential LVDS receivers, this design effectively mitigates the time-walk effect while maintaining the high timing precision required for nuclear and high energy physics research.
The integration of this CFD into the Twin\_Peaks\_CFD1 front-end board demonstrates its practical utility in advanced experimental setups, such as the FATIMA detector at GSI.
Future work could explore adapting the design for a wider range of detector types,
potentially using PCB striplines as delay elements to make the system even more compact.



%
%




\begin{thebibliography}{99}

\bibitem[1]{Gedcke}
D.A. Gedcke, W.J. McDonald,
\emph{A constant fraction of pulse height trigger for optimum time resolution}
Nuclear Instruments and Methods,
Volume 55, 1967, Pages 377-380

\bibitem[2]{jolie}
A. Harter and M. Weinert and L. Knafla and J.-M. Régis and A. Esmaylzadeh and M. Ley and J. Jolie,
\emph{Systematic investigation of time walk and time resolution characteristics of CAEN digitizers V1730 and V1751 for application to fast-timing lifetime measurement},
Nuclear Instruments and Methods in Physics Research Section A,
Volume 1053, August 2023, 168356

\bibitem[3]{mistry}
A. K. Mistry et al.,
\emph{The DESPEC setup for GSI and FAIR}
Nuclear Instruments and Methods in Physics Research Section A,
Volume 1053, August 2023, 168356

\bibitem[4]{fatima}
M. Rudigier et al.,
\emph{FATIMA — FAst TIMing Array for DESPEC at FAIR}
Nuclear Instruments and Methods in Physics Research Section A,
Volume 969, 21 July 2020, 163967











\end{thebibliography}
\end{document}